\documentclass{aa}
\usepackage{graphicx}
\usepackage{rotating}
\usepackage{natbib}
\bibpunct{(}{)}{;}{a}{}{,}

\newcommand{\exo}{EXO 0748--676}
\newcommand{\xmm}{XMM--Newton}

\begin{document}

\title{XMM-Newton light curves of the low-mass X-ray binary \\ EXO
0748--676: dips, eclipses, and bursts}

\titlerunning{ EXO 0748--676: dips, eclipses, and bursts}

\author{Jeroen Homan \inst{1}
   \and Rudy Wijnands \inst{2}
   \and Maureen van den Berg\inst{1}}
   
\institute{INAF/Osservatorio Astronomico di Brera, Via Bianchi 46,
      23807 Merate (LC), Italy; homan@merate.mi.astro.it,
      vdberg@merate.mi.astro.it \and School of Physics and Astronomy,
      University of St Andrews, North Haugh, St Andrews Fife, KY16
      9SS, Scotland, UK; radw@st-andrews.ac.uk}

\offprints{Jeroen Homan, \email{jeroen@space.mit.edu}}

\date{Received / Accepted}
  
\abstract{We present an analysis of \xmm\ light curves of the
dipping, bursting, and eclipsing low-mass X-ray binary \exo, focusing
on the variability on time scales of seconds to hours. The observed
variability can be roughly divided in three types: dips, eclipses and
bursts. We find that the appearance of the latter two, depends
strongly on the strength of the first. We show that the absorption
dips change from spectrally hard to spectrally soft as they become
deeper, which supports suggestions that the source is composed of a
spectrally hard compact source and a spectrally soft extended source.
The fast variability in the soft light curve indicates that the large
structures causing the dips are made up of smaller absorption cores.
We present the first clear detection with \xmm\ of eclipses below 2
keV in this source, and show that dipping activity is apparently
unrelated to the source luminosity. We also test several proposed
models for the spectral evolution during dips and confirm the presence
of a scatter/reflection component in the eclipse spectrum.

 }

\maketitle

\section{Introduction} Among the $\sim$150 known low-mass X-ray
binaries \citep{livava2001}, there are 11 whose light curves show
irregular and sometimes periodic intensity dips. These dips usually
are ascribed to obscuration of the central X-ray source by  thickened
regions in the outer disc, that result from the impact of the
accretion flow from the companion star \citep{whsw1982,wamacl1982}.
Four of the dip sources also show (partial) eclipses, which are due
to obscuration of the central  X-ray source and/or the accretion disc
corona by the companion star. The observed dips and eclipses are
related to the inclination at which those sources are seen, which is
probably between 75$^\circ$ and 90$^\circ$ \citep{frkila1987},
depending on the exact properties of the two stars and the orbital
parameters. Because of their inclination and the regular obscuration
of the bright inner parts of the system, these sources provide a good
opportunity to study the vertical and large scale structure of
accretion flows in low-mass X-ray binaries.

The transient source \exo\ is currently the only active low-mass
X-ray binary that shows both dips and eclipses. Although the source
was discovered in 1985 during an EXOSAT slew \citep{pawhgi1985}, it
was detected in Einstein archival observations from May 1980
\citep{pawhgi1986}; the source has probably remained active ever
since, with 1--20 keV fluxes between $\sim10^{-12}$ erg cm$^{-2}$
s$^{-1}$ \citep[with Einstein;][]{pawhgi1986} and $1.5\,10^{-9}$ erg
cm$^{-2}$ s$^{-1}$ (peak of the 1985 outburst). Observations of type
I X-ray bursts \citep{gohapa1986,gohapa1987} demonstrate that the
compact object is a neutron star. The eclipses and dips revealed an
orbital period of 3.82 hr \citep{pawhgi1986}. Depending on the
stellar parameters, the eclipse duration (8.3 minutes) and the
orbital period give a system inclination of 75--82$^\circ$. With
EXOSAT, dipping activity (up to 80\% decrease in 2--10 keV intensity)
was observed at orbital phases $\sim$0.8--0.2 and $\sim$0.65 (phase
zero being the center of the eclipse), with shallower dips at other
phases.  ASCA observations revealed that at low energies (1--3 keV)
dipping (up to 100\%) is present around most of the orbital cycle
\citep{chbado1998}. 

Observations with the Rossi X-ray Timing Explorer (RXTE) revealed the
presence of quasi-periodic oscillations (QPOs) with frequencies of
$\sim$0.4--3.0 Hz and 695 Hz \citep{hojowi1999,hova2000}. The $\sim$1
Hz QPO was only observed at low luminosities; a similar type of QPO
was also seen in two other high-inclination systems
\citep{jovawi1999,jovaho2000}. This QPO is probably caused by an
opaque structure on top of the accretion disc that orbits the central
source at a distance of $\sim$1000 km, or by variations in the inner
coronal flow  as hydrodynamical simulations of the accretion disc
corona show \citep{mezyfr1991}. The 695 Hz QPO, observed during a
short period of increased luminosity early 1996, had properties
similar to the kHz QPOs observed in many other neutron star sources
\citep{va2000}. 

Using the Reflection Grating Spectrometer (RGS) on board \xmm,
\citet{cokabr2001} recently found emission and absorption features
whose properties also suggest the presence of a thickened accretion
disc with plasma orbiting high above the binary plane. In their
0.3--2.5 keV light curves no eclipses were seen at the expected
times, which was confirmed by the apparent lack of eclipses in the
0.5--2 keV band  of the \xmm\ EPIC-MOS and EPIC-pn observations
analyzed by \citet{bohafe2001}. Based on their results, the latter
authors proposed a model in which the source is a superposition of a
compact ($\sim2\,10^8$ cm) spectrally hard Comptonizing corona and a
more extended ($\sim3\,10^{10}$ cm) spectrally soft thermal halo.
Spectral variations are then caused by variations in the absorption
that affects the hard spectral component more than the extended soft
component. This is in contrast with previous models of the source
consisting of a  compact black body and  extended Comptonized
component \citep{chbado1998} or an absorbed plus un-absorbed
Comptonized component \citep{pawhgi1986}.  

In this paper we present an analysis of \xmm\ light curves, focusing
on variations on time scales of seconds to hours, both at high ($>$2
keV) and low ($<2$ keV) energies.

\section{Observations and analysis}\label{sec:obs}

For our analysis we used all \xmm\ observations of \exo\ that were
publicly available at the time of writing. Although \xmm\
\citep{jalual2001} can observe simultaneously with different
instruments, in many of our observations (all performed during the
calibration and performance verification phase) data from one or more
instruments are absent or incomplete. We decided to use either
EPIC-MOS1 \citep{tuabar2001} or EPIC-pn \citep{stbrde2001} data,
depending on which of these cameras had the longest exposure. A log
of the observations, including the instruments from which the data
were used, is given in Table \ref{tab:log}. Note that the instruments
were used in a variety of modes and with different filters. An
analysis of observations 1 and 2 has previously been published by
\citet{bohafe2001}.

The MOS and PN data were reprocessed following the procedures
described in the \xmm\ data-analysis guides\footnote{The \xmm\ SOC
{\em User's Guide of the \xmm\ Science Analysis System} and {\em An
Introduction to \xmm\ Data Analysis} by the NASA/GSFC \xmm\ GOF} and
using the latest calibration data. Data were extracted from circles
centered around the source, with radii of 90\arcsec\, (MOS1) or
40\arcsec\, (pn, in order to avoid the edges of the CCDs). Light
curves were produced with a time resolution of 1 s and 10 s in the
0.3--10, 0.3--2 and 2--10 keV energy bands. The 10 s light curves are
shown in Fig.\ \ref{fig:curves}. A colour was defined as the ratio of
the count rates in the 2--10 keV and 0.3--10 keV bands, and colour
curves were created by taking the running average of 30 s intervals.
Background light curves were extracted in a similar fashion (same
radius, same CCD),  with the center of the extraction circle at a
location were no source contribution is expected - background
extraction from an annulus centered on the source, as is common
practice, could not be performed, since the source was often located
too close to the edge of the CCD. Variability in the background was
never strong enough to warrant the exclusion of part of our data (see
Fig. \ref{fig:curves}, gray lines). Note that the count rates shown
in Figs.\ \ref{fig:curves}--\ref{fig:lc_fast} are not
background subtracted.

\begin{table}

\caption{Log of \xmm\ observations and data of \exo\ used in this
paper. From left to right: observation number for this paper,
observation date with start and end time, the camera (EPIC-MOS1 or
EPIC-pn) with mode and filter, and exposure time.}\label{tab:log}

\begin{center}
\begin{tabular}{cccc}
\hline
\hline
Obs.\ & Date \& Start-End Time & Camera             & Exp.\  \\
      & (TT)   & [Mode,Filter]$^a$  & (ks)  \\
\hline
1 & 2000-04-04 17:19--22:53 & MOS1 [FF,T] & 20.0 \\
\hline
2 & 2000-04-21 03:59--09:04 & PN [FF,M] & 18.1 \\
  & 2000-04-21 09:50--10:44 & PN [LW,M] & 3.2 \\
  & 2000-04-21 15:22--20:24 & PN [SW,M] & 18.1 \\
\hline
3 & 2001-02-03 17:32--19:32 & MOS1 [FF,M] & 7.2 \\
  & 2001-02-04 01:23--11:54 & MOS1 [FF,M] & 21.6 \\
\hline
4 & \mbox{\hspace{0.15cm}}2001-10-13 00:19--01:01$^b$ & MOS1 [FF,M] & 32.9 \\
\hline
\end{tabular}
\end{center}
$^a$ Mode: FF = Full Frame, LW = Large Window, SW = Small Window;
Filter: T = Thin, M = Medium \\ $^b$ Observation ended on 2001-10-14 
\end{table}

Due to the brightness of the source, significant pile-up occurred in
our observations, especially in the last observation, where a clear
hole could be seen in the center of the point spread function. To
investigate the effects of the pile-up on the shape of light and
colour curves, we made light curves by extracting data only from an
annulus around the source position which excluded the piled-up core
of the point spread function. We found that after scaling to the
original count rates, within the statistics, those light and colour
curves were indistinguishable from those made using the uncorrected
data. In view of our statistics, we decided to use the initial light
curves and not correct for pile-up. Because of the pile-up effects,
which have a more serious effect on spectra than on light curves, we
decided not to perform a detailed spectral analysis, except for the
last the exposure of obs.~2. Details of this analysis, which was
performed with the aim of testing the spectral models of
\citet{chbado1998} and \citet{bohafe2001}, can be found in the
appendix.

\section{Results}\label{sec:results}

\begin{sidewaysfigure*}
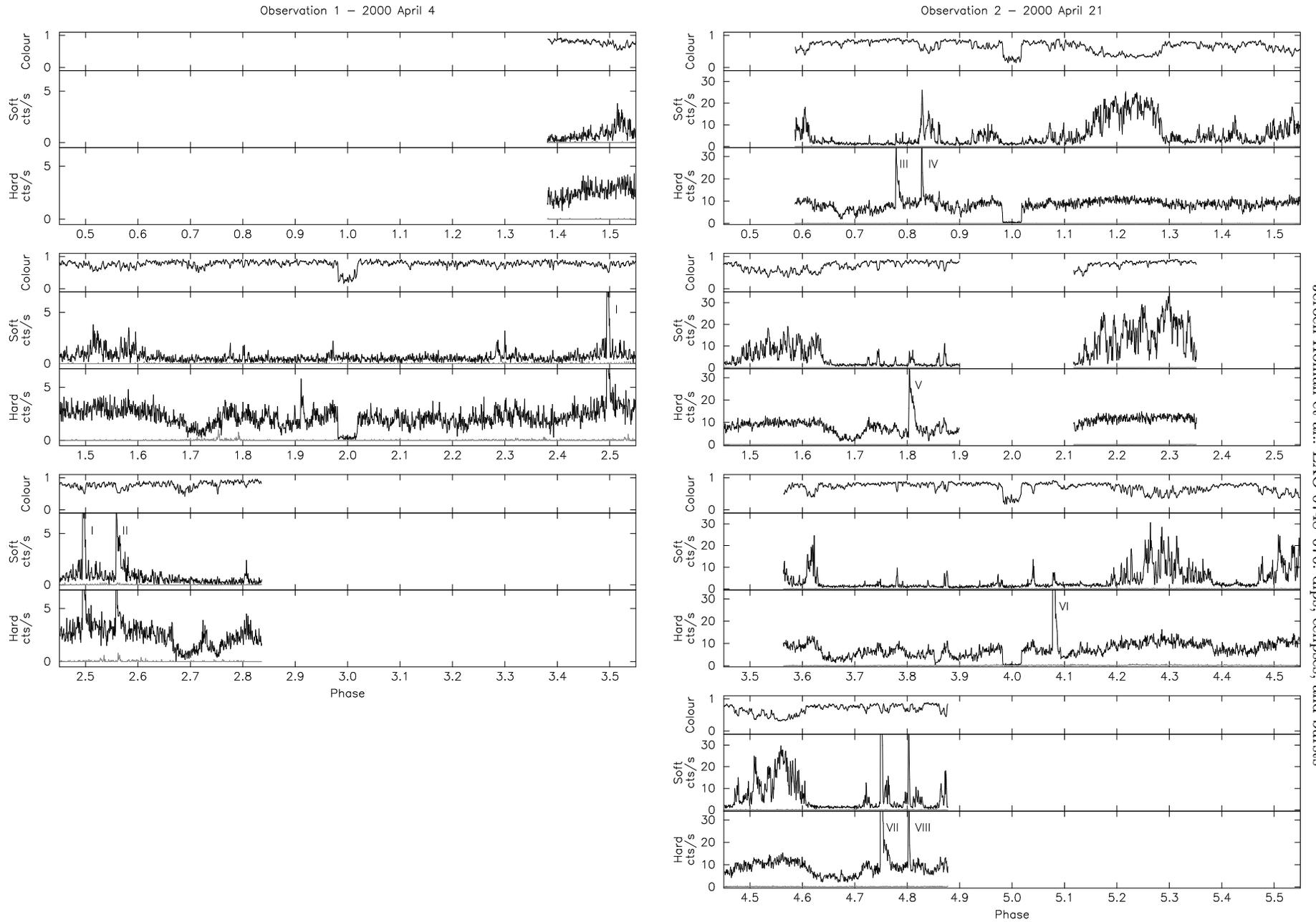
 \hspace{0.5cm}
\includegraphics[height=17cm]{apr4_2000.ps} \hspace{0.5cm}
\includegraphics[height=17cm]{apr21_2000.ps} \hspace{0.5cm}
\caption{\xmm\ light and colour curves of EXO 0748-676 for the 2000
April 4 (a) and 2000 April 21 (b) observations. Each panel shows
roughly one orbital phase, with, from top to bottom, the colour
curve, the 0.3--2 keV light curve, and the 2--10 keV light curve. For
reference we also plot the background light curves (gray lines). The
light curves have a time resolution of 10 s, whereas the colour
curves show a 30 s running average. For clarity, error bars on the
curves were omitted; they are in the order of
$\sqrt{0.1\times\textup{count rate}}$. Type I X-ray bursts are
indicated by Roman numerals.} \label{fig:curves}
\end{sidewaysfigure*}

\addtocounter{figure}{-1}

\begin{sidewaysfigure*}
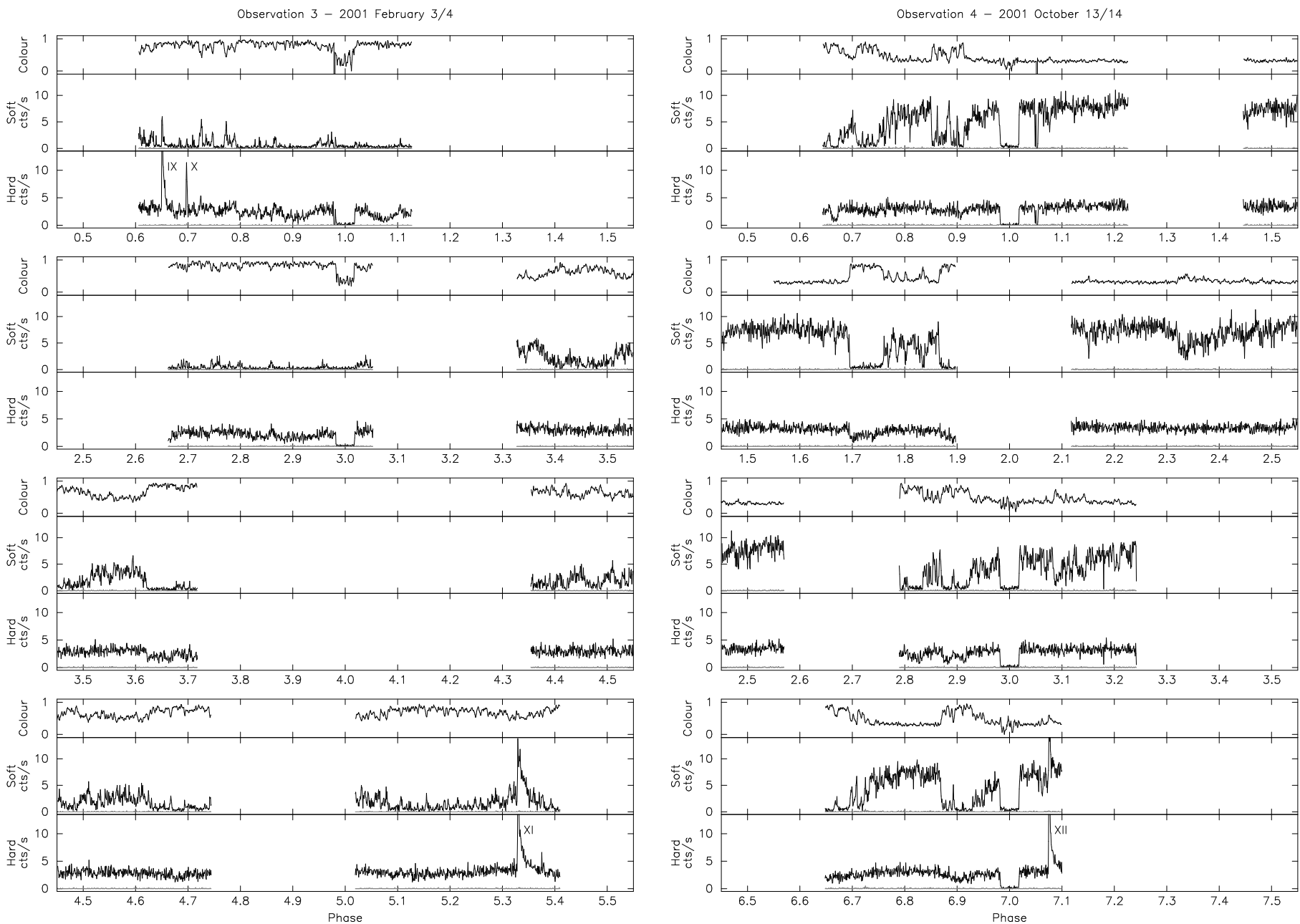
 \hspace{0.5cm}
\includegraphics[height=17cm]{feb_2001.ps} \hspace{0.5cm}
\includegraphics[height=17cm]{oct_2001.ps} \hspace{0.5cm}
\caption{\ldots continued. Light and colour curves for the 2001 Feb
3/4 (c) and 2001 October 13/14 (d) observations. The sharp dips in
the first orbits of panels c (just before the eclipse ingress) and d
(around phase 1.05), during which no counts were observed, are
probably instrumental effects.} \end{sidewaysfigure*}

\begin{figure}[t] 
\resizebox{\hsize}{!}{\includegraphics{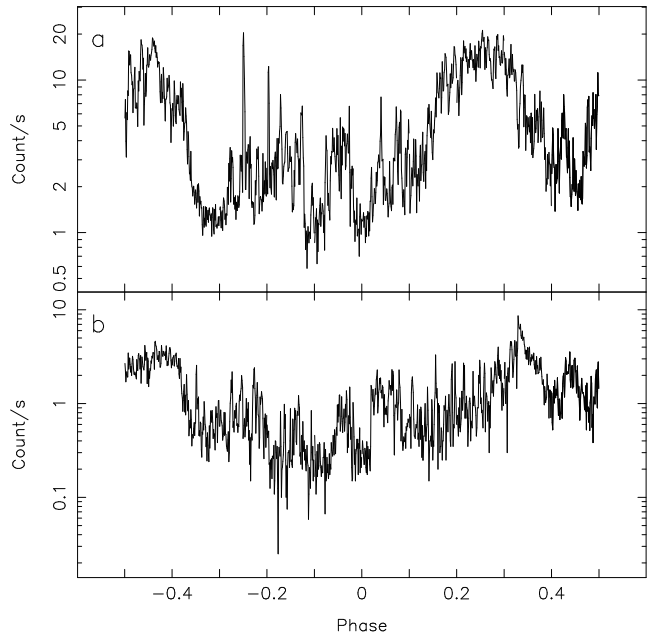}}
\caption{Folded 0.3--2 keV light curves of observation 2 (a) and
observation 3 (b). The data were rebinned to 1000 bins per orbital
cycle and, unlike Fig.~\ref{fig:curves}, the count rate is plotted
logarithmically. Type I bursts are still included. Weak eclipses are
visible around phase zero. }\label{fig:fold} \end{figure}

In Figs. \ref{fig:curves}a--d we show the light curves of \exo\ in
the 0.3--2 keV (soft) and 2--10 keV (hard) energy bands, together
with curves for the colour. Count rates are plotted versus orbital
phase, with phase zero assigned to mid-eclipse. Note that, because of
different instruments, filters, and CCDs, the count rates of the four
observations should not be compared directly; the low count rates in
observation 1 are due  to the use of the thick filter instead of the
medium filter, whereas the high count rates of observation 2 are due
to the higher sensitivity of the EPIC-pn detectors. 

As was already noted by \citet{bohafe2001}, at first sight the soft
and the hard light curves have little in common; the soft light
curves show a lot of rapid activity, whereas the variations in the
hard light curves,  except for eclipses and type I X-ray bursts, are
much slower and usually weaker. 

\subsection{Eclipses}

\begin{figure}[t] 
\resizebox{\hsize}{!}{\includegraphics[angle=-90]{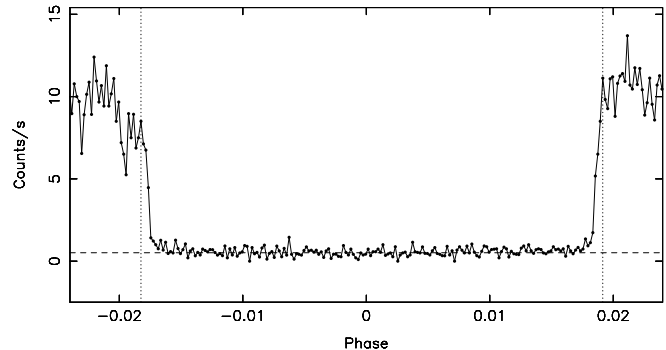}}
\caption{An enlargement of part of the folded 0.3--10 keV light curve
of observation 4. The time resolution is 2 s. The dashed line
represent the average count rate from 200 data bins centered around
phase zero. The dotted lines represent the start of eclipse ingress
and the end of eclipse egress.}\label{fig:eclips} \end{figure}

While the RGS and EPIC data analyzed previously \citep{cokabr2001,
bohafe2001} were characterized by the absence of eclipses at low ($<$
2 keV) energies, our fourth observation (Fig.\ \ref{fig:curves}d) is
the first \xmm\ light curve of \exo\ to show clear eclipses at low
energies. In contrast to those in the soft band, the light curves in
the hard band always show eclipses. It should be noted that in the
folded light curves of observations 2 and 3 (shown in
Fig.~\ref{fig:fold}) during the phases of the eclipse the count rates
in the soft band are actually below the average count rates for these
observations (but not as low as some of the dips). Although not as
apparent as in observation 4, this probably indicates that in these
observations the effects of the eclipse {\it are} present. In both
bands the eclipses are not total, with residual count rates of
$\sim$5--8\% and $\sim$4--5.5\% in, respectively, the soft and hard
band of the fourth observation. Typical ingress/egress times in
observation 4 are $\sim$8--10 s (see Fig.~\ref{fig:eclips}), both in
the soft and hard band. The folded 0.3--10 keV light curve of
observation 4 (Fig.~\ref{fig:eclips}) also reveals the presence of
structures at the end of the eclipse ingress and beginning if the
eclipse egress, similar to the 'shoulders' reported by
\citet{pawhgi1985}.

\subsection{Dips}\label{sec:dips}

\begin{figure}[t]
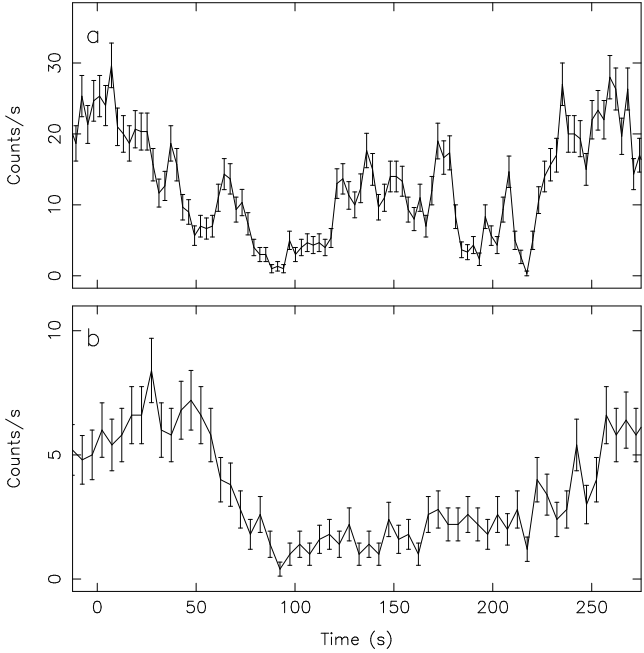
  \resizebox{\hsize}{!}{\includegraphics{fast_a.ps}}
\resizebox{\hsize}{!}{\includegraphics{fast_b.ps}} \caption{Light
curves showing the fastest observed dipping events in the 0.3--2 keV
(a) and 2--10 keV (b) bands. The 0.3--2 keV data are taken from
observation 2 around phase 2.2 and have a time resolution of 3 s; the
2--10 keV are taken from the same observation around phase 3.85 and
have a time resolution of 5 s.}\label{fig:lc_fast} \end{figure}

Like in the ASCA observations analyzed by \citet{chbado1998}, dipping
extends around most of the orbital phase and is most clearly visible
in the soft band - only observation 4 shows several intervals in
which the 0.3--2 keV light curve is dip-free. With dipping being
present most of the time, the most eye-catching features in the soft
band are periods during which the strength of the dipping is reduced.
These decreases in the dipping are the soft flares reported by
\citet{cokabr2001} and \citet{bohafe2001}. Periods of reduced dipping
show two typical time scales; extended intervals, lasting typically
for several minutes to an hour (Fig.~\ref{fig:curves}) and shorter
changes during those intervals, with time scales of seconds
(Fig.~\ref{fig:lc_fast}a). When present, the extended intervals of
(reduced) dipping in the soft and hard bands are probably related to
the orbital phase; they appear at a relatively constant phase and are
present for several consecutive orbits. This is most clearly seen in
observation 2. Note that the interval of reduced dipping around phase
0.25 in this observation seems to show a slow, possibly random, shift
towards higher phases; taking the moment during ingress at which the
'saturation' level in the soft band is reached as a reference point,
the ingress of the dipping interval at phases near 0.3, seems to
occur  at a later ($\sim$3\%) phase in subsequent orbits.

The phase extent and strength of the (decreases in) dipping undergo
changes between orbital cycles. Unlike the long  intervals, smaller
structures do not seem to recur in later orbits - at least not at a
similar phase.  The rapid variations in the count rate in the soft
band can be intense, with the count rate changing by factors of more
than 10 on time scales of only a few seconds (see e.g.
Fig.~\ref{fig:lc_fast}a) - they are most clearly observed in
observations 2 and 4. Excluding type I X-ray bursts and eclipses, the
dynamical range (defined as the maximum count rate divided by minimum
count rate, within one observation) of the soft band is $>$110.
Variations in the hard band, seem to be more modest; the dynamical
range is only $\sim$30, and strong/violent short-time variations,
like those seen in the soft band, are not observed. For comparison we
show the fastest observed dips in the soft and hard bands in
Fig.~\ref{fig:lc_fast}; with a dip ingress of more than 30 seconds
the fastest dip in the hard band is considerably slower than those in
the soft band, which can be as fast a few seconds. During the low
count rate periods (including the eclipses) the source is detected
well above the background in the 0.3--2 keV band, with minima
(observations 3 and 4) of $\sim$0.1--0.3 cts s$^{-1}$, compared to
background count rates of $\sim$0.01--0.03 cts s$^{-1}$.  

\begin{figure}[t] \resizebox{\hsize}{!}{\includegraphics{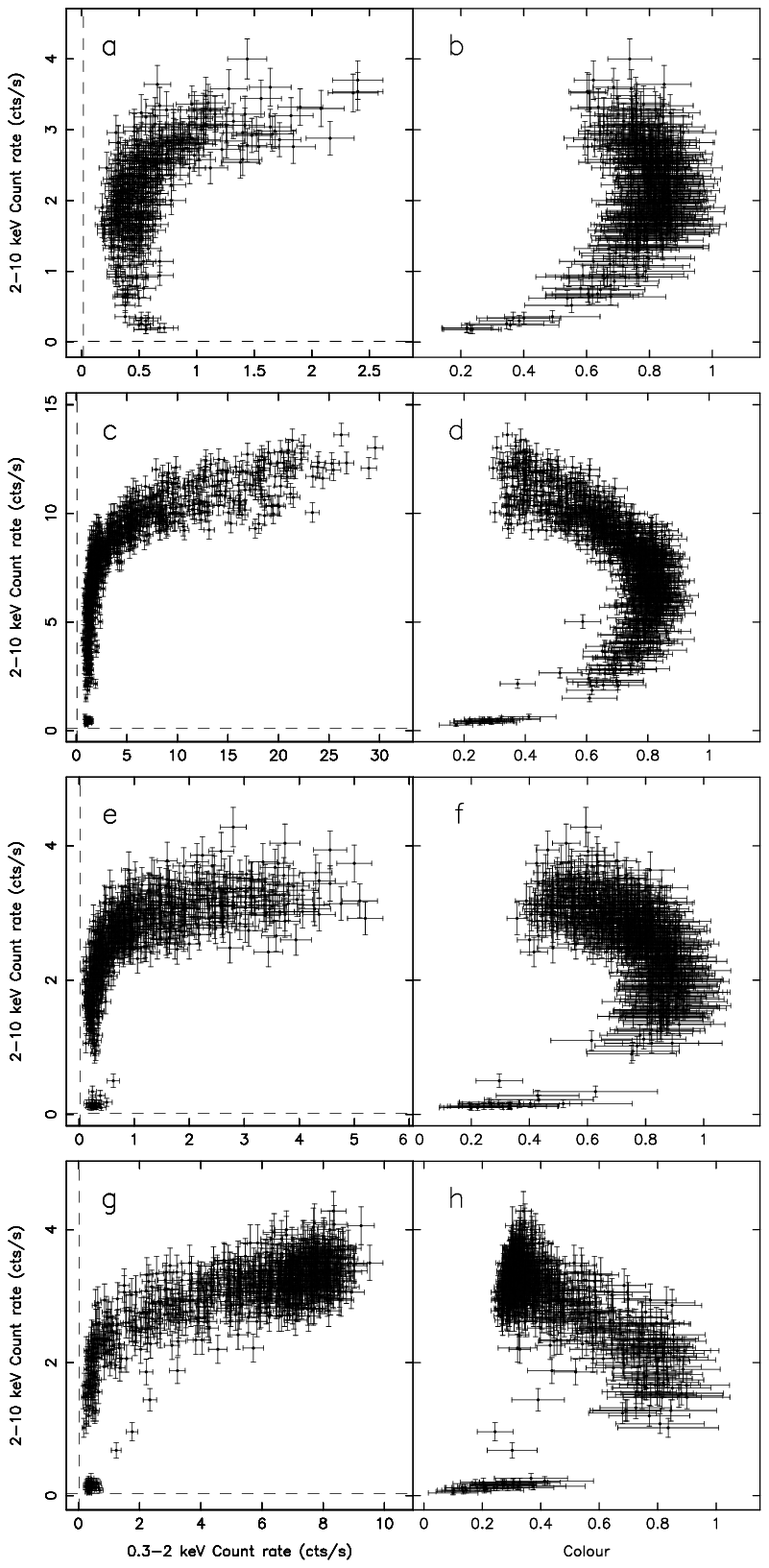}}
\caption{Left panels: 2--10 keV count rate vs.\ 0.3--2 keV count
rate, Right panels: 2--10 keV count rate vs.\ colour. From top to
bottom: observation 1, 2, 3, and 4.  Each point represents a 50 s
average. The two dashed lines in the left panels represent the
average background count rates in the two energy bands.  The eclipses
can be seen as small detached blobs. Note that while the count rates
in the soft band during the eclipse are similar to those during the
dips, they are clearly lower for the hard band. }\label{fig:soft-hard}
\end{figure}

As expected, the hard band is less affected by the dipping than the
soft band. Dips in the hard band are only (but not always) observed
when the count rate in the soft band is at a minimum.  They are
mostly observed at the phases around 0.7 and 0.9, which are also the
phases which show the least variability and the lowest count rates in
the soft band. The relation between dipping in the soft and hard band
is clarified when we plot the count rates in the hard and soft band
against each other, as has been done in Fig.\ \ref{fig:soft-hard}; 
high count rates in the soft band are only found when the count rate
in the hard band is high (i.e.\ outside dips).  Dips in the soft
count rate are however not always visible as dips in the hard band. 
During dips in the hard band the count rate does not drop below the
level found in eclipses, whereas in the soft band the source can
occasionally be weaker than during eclipse (see also
Fig.~\ref{fig:fold}). In all four observations the spectrum initially
hardens as the count rate decreases, until the hard band starts to be
affected, after which the spectrum starts to soften (see also Fig.~\ref{fig:spectra}). It is also
interesting to see that in observation 4 the colour saturates at the
highest count rates and that the spectrum during the eclipse is
softer than this saturation level. The motion of the source along the
tracks in Fig.~\ref{fig:soft-hard} is smooth, in the sense that
(except during eclipses) the source never jumps from one end of the
track to the other.

\begin{figure}[t]
\resizebox{\hsize}{!}{\includegraphics{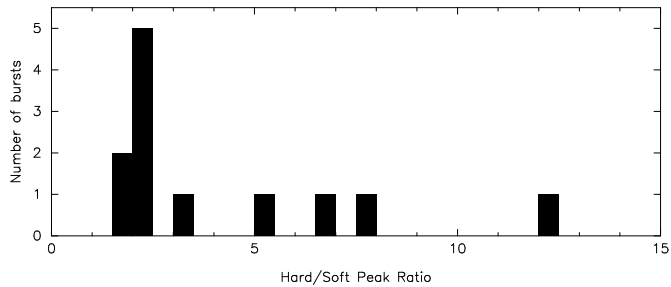}}
\caption{Histogram showing the observed ratio of peak count rates of
the hard and soft bands for the 12 observed type I X-ray bursts.
See also Table \ref{tab:burst}} \label{fig:bursts} \end{figure}

\begin{table} \caption{Peak count rates of the 12 observed type I
X-ray bursts in the soft (0.3--2 keV) and hard (2--10 keV) X-ray
bands, together with their hard/soft ratio.  Indicated is whether the
bursts occurred during intervals of reduced dipping or full dipping
(corresponding to the lowest count rates in the soft band, and, as
opposed to reduced dips, sometimes also visible as dips in the hard
band. For bursts I, II, and VII this was not completely clear. See
Figure \ref{fig:curves} for the labeling (I--XII) of the
bursts.}\label{tab:burst} \begin{center}
\begin{tabular}{lllll}
\hline
\hline
Burst & Pers.$^a$ & Soft peak & Hard peak & Ratio \\
\hline
I    & R(?) & 9.8$\pm$1.4  & 19.8$\pm$2.0	& 2.0$\pm$0.4 \\
II   & R(?) & 8.8$\pm$1.3  & 18.4$\pm$1.9	& 2.1$\pm$0.4 \\
III  & F & 8.4$\pm$1.3     & 42.0$\pm$2.9	& 5.0$\pm$0.8 \\
IV   & R & 20.6$\pm$2.0    & 43.0$\pm$2.9	& 2.1$\pm$0.3 \\
V    & F & 5.0$\pm$1.0     & 38.0$\pm$2.8 	& 7.6$\pm$1.6 \\
VI   & F & 8.6$\pm$1.3     & 104$\pm$5  	& 12.1$\pm$1.9 \\
VII  & R(?) & 87.4$\pm$4.2 & 162$\pm$6	        & 1.9$\pm$0.1  \\
VIII & R & 56.0$\pm$3.3    & 99.4$\pm$4.5	& 1.8$\pm$0.1 \\
IX   & F & 8.0$\pm$1.3     & 27.0$\pm$2.3	& 3.4$\pm$0.6 \\
X    & F & 1.8$\pm$0.6     & 12.4$\pm$1.6  	& 6.9$\pm$2.5 \\
XI   & R & 10.8$\pm$0.5    & 24.4$\pm$2.2   	& 2.3$\pm$0.2 \\
XII  & R & 15.0$\pm$1.7    & 30.4$\pm$2.5   	& 2.0$\pm$0.3 \\
\hline
\end{tabular}
\end{center}
\noindent $^a$ Type of persistent emission in the soft band during
which the burst occurred; R=Reduced dipping, F=Full dipping \\
\end{table}

\subsection{Type I X-ray bursts}
	   
In total 12 type I X-ray bursts were observed, 8 of which occurred
in pairs with a separation of $\sim$10.65--14.65 minutes.  When
comparing the X-ray bursts in the hard and soft band we find that
some of them hardly show up in the soft band. To illustrate this, we
give the ratios of the hard peak and soft peak count rates for all 12
bursts in Table \ref{tab:burst} and plot them in Figure
\ref{fig:bursts}. Count rates were obtained from 5 s running averages
to reduce the scatter in the 1 s light curves. As no correlation was
found between total count rate and the ratio of the peak count rates
in the two bands, we assume that the effects of pile-up on our
results are negligible. The peaks in the hard and soft band  did not
occur simultaneously - the quoted count rates are those at the time
of the hard peak. Taking the soft count rates from the peak in the
soft band (which occurred between 9 s before and 7 s after that in
the hard band) results in a similar picture. A clear concentration (7
out of 12) of bursts is present around a hard/soft peak ratio of
$\sim$2, with the other bursts all having larger values. While due to
the rapid nature of the dips, it is not possible to rule out that
some of the burst occurred during short periods of reduced dipping
(which have similar time scales as the bursts), it seems that the 5
bursts that have ratios larger than 3 all occur during dipping
intervals. Note that the small peak around phase 1.9 in the hard band
of observation 1 was not included, since the nature of this peak was
not clear. If this peak, which occurred during a dip in the hard
band, also was a type I burst its hard/soft ratio would be even
higher than that for burst VI.

\subsection{Fluxes}

In order to compare the fluxes of the four observations as well as
possible, we fitted the spectra of the source during the parts of the
observations with the highest count rates. The spectra were fitted in
{\tt XSPEC} with various models. While most models provided good fits
in the 5--10 keV range, we were not able to satisfactorily fit the
low energy parts of the spectrum with a consistent model (see also Appendix). For all
four observations the unabsorbed 5--10 keV flux was consistent with
$9\,10^{-11}$ erg\,cm$^{-2}$\,s$^{-1}$. Our fits suggest that
differences at lower energies are likely due to the complex effects
of absorption.

\section{Discussion}\label{sec:disc}

While at first sight the count-rate variations at low and  high
energies in EXO 0748--676 seem to occur independently from each other
(Fig.\ 1), they are in fact strongly related (Fig.\ 2) and likely due
to the same mechanism. This is supported by the fact that the source
moves smoothly along the track in Fig.\ 2, which suggests that most
of the variations in the light curve are caused by changes in only
one parameter. In this section we discuss how changes in the
absorption can account for the observed variations as well as for the
appearance of the eclipses and type I X-ray bursts. 

\subsection{Dips, eclipses, and bursts}

Both \citet{chbado1998} (ASCA observations) and \citet{bohafe2001}
(\xmm\ observations) showed that the spectral evolution of \exo\ can
be largely explained by the effects of variable absorption
progressively covering two (stable) emission components. In both
models the extended component is spectrally softer than the compact
central source and as such both are consistent with the observed
behavior. However, as we show in the Appendix, the real spectral
composition is likely to be more complex than suggested by
\citet{chbado1998} or \citet{bohafe2001}.

As absorption is intrinsically less effective at high energies,
dipping is most pronounced at low energies and therefore spectrally
hard. This can clearly be seen in our observations 1--3 and in the
ASCA light curves of \citet{chbado1998}, in which dipping in the low
energy bands extends over a much larger phase than in the high energy
bands; the  intervals of rapid increases in the count rate in
observations 1--3 and most of observation 4 represent periods in
which the absorption is reduced by a significant fraction.  One would
expect the ingress of dips in the soft band to precede that in the
hard band, an effect that is indeed seen in our light curves and even
more clearly demonstrated in Fig.\ \ref{fig:soft-hard}. We therefore
prefer to refer to the intervals of increased count rate as periods
of reduced dipping instead of flaring, since we think this gives a
better description of what gives rise to the changes in the light
curves. 

\citet{bohafe2001} suggested that part of the variability in their
'flares' is due to intrinsic changes in the halo luminosity (possibly
due to reprocessing of type I X-ray bursts). However, the fact that
the count rates in the eclipse in observation 4 are similar to those
in observation 3, which had the same instrumental setup but was much
less luminous outside the dips in the soft band, suggests (assuming
that the same part of the halo was left uncovered during the eclipse)
that the halo luminosity did in fact not vary significantly.
Moreover, since the 'flares' occur at similar phases for several
orbits, delayed reprocessing of type I X-ray bursts is basically
ruled out as a trigger for halo-luminosity changes.

The apparent lack of eclipses in the soft band of the first three
observations can easily be explained; the additional absorption
column presented by the companion star has almost no influence below
2 keV, since most of the soft flux from the central source was
already absorbed at the time of eclipse - the projected size of the
halo also has to be significantly larger than that of the companion
star to not be affected during an eclipse.  The fact that the count
rates outside of the eclipse are sometimes lower than those during
the eclipse suggests that the absorbing structures, as seen from the
central source, occasionally subtend a larger solid angle than the
companion star. This indicates that in the vertical direction the
absorbing material extends well beyond the expected scale height of a
thin disc. 

The behavior of the type I X-ray bursts can also be explained in this
framework; when they occur during intense dipping intervals, i.e.
when absorption is high, the soft flux of the burst is more affected
than the hard flux, resulting in ratios larger than 2. Those that
occur during intervals of reduced dipping are affected less by
absorption, resulting in ratios of $\sim$2, which is apparently the
normal ratio for these bursts in this source. The spread in ratios
larger than 2 probably reflects various stages of absorption.  We
note that many of the bursts analyzed by \citep{copame2002}, which
showed red-shifted absorption lines, occurred during periods of
significant absorption below 2 keV.

\subsection{Location of the absorbing material}

Since we know that the absorbing material has a height of more than
8--15$^\circ$ above the orbital plane it has to be constantly
refreshed in order to be visible at a more or less fixed position in
the orbital frame, for at least several orbital cycles; if not, it
would disappear on a time scale that is shorter than the orbital
period, because the time scale for changes in the vertical scale
height in the accretion disc is comparable to the dynamical time
scale \citep{frkira1992}. Building on the work of \citet{lush1976},
\citet{frkila1987} proposed a model in which an accretion stream
overshooting the point of impact with the disc settles in a ring-like
structure at a radius of $\sim10^{10}$ cm (for a 1.4 $M_\odot$
neutron star and $P_{orb}=3.82$ hr). As the result of an ionization
instability, a two-phase medium is formed, which consists of a hot
thin gas and cool clouds and is present between phases 0.3 (at which
the material joins the disc) and 0.8. The presence of such a
two-phase absorber, consisting of both neutral and ionized gas, seems
to be confirmed by recent Chandra observations of the source
\citep{jiscma2003}. If we apply the equations of \citet{frkila1987}
to EXO 0748-676, we find a typical size for the cool clouds of
$4\,10^7$ cm and a column density of $6\,10^{23}$ cm$^{-2}$. At a
distance of 10$^{10}$ cm this should cause dips with a time scale of
$\sim$1 s, quite similar to the shortest observed time scales. While
these clouds can explain the rapid  variations in the soft band, they
cannot (within this model) explain the occurrence of dipping at
phases between 0.8 and 0.3.  

Another possible site of absorbing structures could be a thickened
disc rim; \citet{rechbe2002} found that the dips in the
high-inclination X-ray binary X1916--053 are responsible for the
positive superhump period observed in the X-rays and optical. They
suggest that a thickened disc rim can explain both the X-ray and
optical superhump. Positive superhumps (quasi-periodicities a few
percent different from the orbital period) are observed in some
cataclysmic variables \citep{wa1995} and low-mass X-ray binaries
\citep{odch1996} and are interpreted as a signature of an eccentric
precessing accretion disc. Evidence for such superhumps may be
present in observation 2 in the form of dipping intervals shifting in
phase (see section \ref{sec:dips}). However, the phase shift of these
dips might as well be a random effect; unfortunately, the data sets
for the other observations either are too short or contain too many
gaps to study this in more detail. 

Whatever the exact nature of the absorbing material is, the fourth
observation shows that its structure and/or presence can change
considerably on time scales shorter than eight months (which is the
interval between observations 3 and 4). It is not clear what causes
these changes. The 5--10 keV flux during the brightest intervals of
each observation are consistent with each other, suggesting that, if
proportional to the 5--10 keV flux, the mass accretion rate remained
constant. A possible explanation could be the precession of a tilted
accretion disc, evidence for which has been found in several other
low-mass X-ray binaries \citep[e.g.][and references therein]{la1998}.
A tilted accretion disk, and hence precession, may also be present in
\exo\ as was suggested by \citet{crsthu1986} on the basis of optical
observations. If true, the absorbing structures could remain
unchanged, but depending on the phase of the precession more (or
less) of the absorbing material moves through our line of sight.

\subsection{Constraints from time scales}

The ingress and egress time scales we observed during the eclipses
are on the order of $\sim$8-10 s, somewhat larger but consistent with
values obtained with EXOSAT and RXTE \citep{pawhgi1986,hewoco1997}.
Depending on the density profile of the companion star's outer
atmosphere \citep[see a discussion on this topic in][]{pawhgi1986},
this means that 92--95\% of the 0.3--10 keV flux and 94.5--96\% of
the 2--10 keV flux originates in a region with a radius of
$3.5-3.9\,10^8$ cm or smaller. The latter number is obtained from the
duration of the eclipse ingress and egress; depending on the
properties of the secondary \citep[see][]{pawhgi1986}, binary
parameters (such as secondary's radius, binary separation and inclination) were derived, which, using the eclipse time scales,
translated into upper limits on the size of the eclipsed central
object. As was already found by \citet{pawhgi1986}, near the bottom
level of the eclipse, the eclipse ingress and egress progress slower
than at the higher count rate levels (Fig.~\ref{fig:eclips}). This
cannot be due to the density profile of the secondary's atmosphere,
and suggests a central source geometry that is at least partly
extended. 

The shortest time scales in the soft band are in the order of a few
seconds and  those in the hard band are $\sim$30 s. These time scales
are difficult to interpret, since the distance from the central
source and velocity across our line of sight of the absorbing
material are not well known. However, the time scales for the dips in
the soft band, which are shorter than the eclipse ingress and egress
times, suggest that the central region is smaller than the values
given above.  The shortest observed time scales for the two bands are
quite different. This difference can be explained if we assume that
the absorbing material consists of small clouds (patchy absorber) as
in the model of \citet{frkila1987} discussed above, each with a
density high enough to considerably affect the soft flux, but not the
hard flux. As the absorbing material (containing the small clouds)
starts moving into our line of sight, the soft band is immediately
affected, but it takes more time to have the required number of
clouds present in our line of sight to affect the hard flux. An
additional effect of such a patchy absorber is that the variations in
the soft band are much stronger than in the hard band - quite the
opposite of what is found for other types of variability in X-ray
binaries, which usually get stronger towards higher energies.

\subsection{Residual emission}

From Figs.\ \ref{fig:curves} and \ref{fig:soft-hard} it is clear that
eclipses and dips in the soft and hard band both tend to saturate at
a level that is significantly above the background. As discussed
above, part of this residual flux can be contributed to (fluorescent)
emission from the extended thermal halo
\citep{cokabr2001,bohafe2001}. However, as suggested by observations
of type I X-ray bursts during eclipses in \exo\
\citep{pawhgi1985,gohapa1987}, $\sim$4\% of the 2--6 keV flux from
the central source is scattered into our line of sight during
eclipse. This scattering occurs in the accretion disc corona, which
might be the same component as the extended thermal halo, and accounts
for most of the residual flux in the hard band; residual emission in
the hard band during eclipses in observation 4 is also $\sim$4\%. The
slightly higher fraction of residual flux in the soft band (compared
to the hard band) can be contributed to additional emission from the
extended thermal plasma present above the disc.

\section{Summary}

We have analyzed XMM-Newton light curves of the low-mass X-ray binary
EXO0748-676. We find that the appearance of the light curves in the
soft and hard band can be well explained by progressive absorption of
two spectral components. From the colour of the residual emission
during eclipses, and from the colour evolution with increasing degree
of absorption, we conclude that the source contains an extended
source of X-ray emission that is spectrally softer than the compact
source, as suggested earlier by \citet{chbado1998} and
\citet{bohafe2001}. 

In the soft band, the count rate during dipping intervals sometimes
decreases below the level during eclipses, while the source is always
detected well above the background, both in the soft and hard band.
This implies that (at certain phases) the absorbing material as seen
from the central source, subtends a larger solid angle than the
projection of the companion star, but that it is smaller than the
extended soft component. The recurrence of dipping structures in the
light curves in consecutive orbits indicates that the absorbing
material can maintain a more or less fixed position in the corotating
frame for at least several orbits. However, the drastic change of
appearance in the light curve of October 2001 with respect to that of
February 2001, indicates that the position and/or density of the
absorbing material can vary within months.  The difference in the
shortest time scales for dipping activity in the soft and hard band,
suggests that the dipping is caused by a patchy absorber that affects
soft energies more efficiently.

Future observations with \xmm, in particular for several continuous
orbital periods, may shed more light on the possible presence of
superhumps and on the evolution and structure of the absorbing
material.

\begin{acknowledgements}

JH acknowledges support from Cofin grants 2001/MM02C71842 and
2001/028773. MvdB acknowledges financial support from the Italian
Space Agency and MIUR. This work is based on observations obtained
with XMM-Newton, an ESA science mission with instruments and
contributions directly funded by ESA member states and the USA
(NASA).

\end{acknowledgements}


\appendix

\section{Spectral analysis}

Several attempts have been made in the past to model the spectral
evolution of \exo\ as a function of increasing absorption. In this
appendix we test the models by \citet{chbado1998} and
\citet{bohafe2001} that were already discussed extensively in the
main text. Although the soft-band light curves presented earlier in
this paper and the recent spectral results by \citet{jiscma2003}
strongly suggests that the obscuring medium is far from being 
uniform, these models treat the absorber as having a smooth density
profile. Moreover, the extended components that are part of the two
models are assumed to have a uniform brightness. Notwithstanding
these simplifications we feel that a vigorous test of the two models
may yield important information on the structure of the central
source and the absorber.

\subsection{Data reduction}

\begin{figure}[t]
\resizebox{\hsize}{!}{\includegraphics[angle=-90]{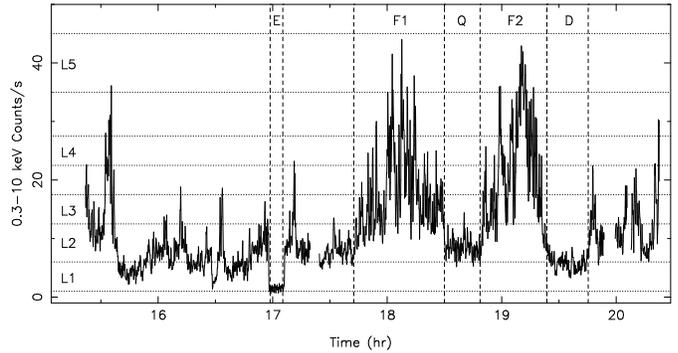}}
\caption{Light curve of the third data segment of obs.~2 (0.3--10 keV)
showing the time and count rate selections applied for our spectral
analysis. Type I X-ray bursts are not shown. The labels L1-L5
indicate the count rate level selections (see main text). The other
labels indicate the time selections and stand for eclipse (E),
flaring interval 1 (F1), quiescence (Q), flaring interval 2 (F2), and
dip (D). Note that the eclipse was not included in the L1 count rate
selection.} \label{fig:select} \end{figure}

For our spectral analysis we used the small-window mode EPIC-pn data
of obs.~2 (i.e. the third data segment), the only part of our data
set that was not affected by pile-up. Source data were extracted from
a circle with a 38\arcsec\ radius, centered around the source;
background data from a similar circle in the opposite corner of the
window. Only events with pattern 0-4 were selected and data from
bursts were discarded. Data were selected either on time or the
0.3--10 keV count rate; the times and count rates for the various
selections can be found and seen in Table \ref{tab:select} and
Fig.~\ref{fig:select}. Note that our time selections include those
made by \citet{bohafe2001} (F1, Q, F2, D) with the addition of an
eclipse spectrum (E). The count rate selections, which do not cover
the entire count rate range (see gaps in Fig.~\ref{fig:select}) were
converted to time intervals using a 10s 0.3--10 keV light curve. The
resulting spectra were fitted with the  Xspec package \citep[V11.2]{ar1996} with the
models described below, using a canned response matrix. The ancillary
response file was created using {\tt arfgen}.

Before we discuss our spectral fits, some comments should be made on
the two selection methods. First, selecting spectra based on temporal
behavior (e.g. dipping activity), as was done by \citet{bohafe2001}
(see selections F1 and F2), will result in mixing of spectra with
rather different properties; as is clearly illustrated by
Fig.~\ref{fig:soft-hard}, spectral proprieties can change
considerably with count rate. Selecting (narrow) count rate
intervals, as was done by \citet{chbado1998}, will reduce these
problems a bit, but mixing still occurs in those intervals where the
count rate does not correlate with colour.  Second, from other
low-mass X-ray binaries we know that spectral evolution of the
central source can occur on time scales that are considerably shorter
than the exposure we analyzed (i.e. a few hours). Spectra that were
taken more than a few hours apart, and which therefore have likely
different central source properties, could end up having similar
count rates because of the varying absorption, resulting in the
mixing of different spectral properties. Unfortunately, tracing
changes in the central source spectrum would require splitting our
exposure in shorter pieces, resulting in a low statistical quality of
the spectra.

\begin{table}

\caption{Time and
count rate selections used for our spectral analysis. These
selections are also displayed in Fig.~\ref{fig:select}. Note that the
eclipse (E) is excluded from the L1 count rate
selection.}\label{tab:select}\begin{center}
\begin{tabular}{ccc}
\hline
\hline
Selection & Time$^a$ & 0.3--10 keV count rate \\
          &  (TT)    & (cts/s) \\
\hline
E         & 16:58:40--17:05:20  & --- \\
F1        & 17:42:30--18:30:00  & --- \\
Q         & 18:30:00--18:48:40  & --- \\
F2        & 18:48:40--19:23:40  & --- \\
D         & 19:23:40--19:45:20  & --- \\
\hline
L1        & ---                 & 1.0--6.0 \\
L2        & ---                 & 6.0--12.5 \\
L3        & ---                 & 12.5--17.5 \\
L4        & ---                 & 22.5--27.5 \\
L5        & ---                 & 25.0--45.0 \\
\hline
\end{tabular}
\end{center}
\noindent $^a$ Time on April 21 2000.  
\end{table}

\begin{figure}[t]
\resizebox{\hsize}{!}{\includegraphics[angle=-90]{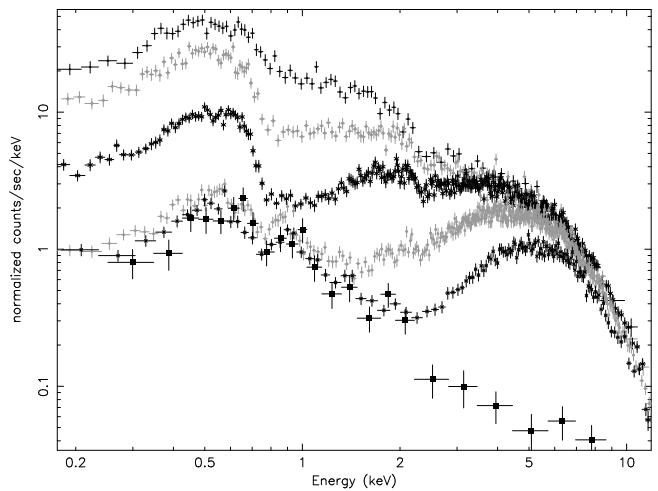}}
\caption{Energy spectra for the 5 count rate selections and the
eclipse interval. From top to bottom: the spectra for the selections
L5 to L1, and the eclipse spectrum (big filled square symbols). All
spectra are rebinned for display purposes.} \label{fig:spectra} \end{figure}

\subsection{Spectral fits} 

The eclipse spectrum and the count rate selected spectra are shown in
Fig.~\ref{fig:spectra}. For a display of the time selected spectra we
refer to Fig.~4 of \citet{bohafe2001}. Fig.~\ref{fig:spectra} clearly
shows that, when going to lower count rate selections (from L5 to L3)
the spectrum above $\sim$3 keV remains largely unchanged, explaining
the hardening of the spectrum (see Fig.~\ref{fig:soft-hard}). From L3
to L1 that part of the spectrum also starts to become affected, which,
because of the saturation at low energies (compare L2 and L1
spectra), leads to a softening of the spectrum. The eclipse spectrum
nicely shows that the soft part of the spectrum is consistent with
that of the deepest dips, whereas at high energies the source is
almost completely obscured (compare eclipse and L1 spectra). 

\begin{table*}[!t]

\caption{Spectral fit results. Models and energy ranges are given for
each of the fits. Lower limits on {\tt ABB} are 90\% confidence. All
other errors represent 1$\sigma$ confidence
intervals.}\label{tab:fits-church}

\begin{center}
\begin{tabular}{llllllll}
 
\hline \hline
 & \multicolumn{3}{l}{Model: Church et al.\,(1998)} & & \multicolumn{3}{l}{Model: Church et al.\,(1998)  + Gaussian} \\
 & \multicolumn{3}{l}{Range: 0.8--10 keV \mbox{~~~} $\chi^2_{red}$/d.o.f.=1.01/1594} & & \multicolumn{3}{l}{Range: 0.15--12 keV \mbox{~~~}
$\chi^2_{red}$/d.o.f.=1.18/1916} \\
\hline
                      & L5 & L3 & L1 & & L5 & L3 & L1 \\
\hline
{\tt ABB} ($^{a}$)    & 0.01(19) & 0.2(2) & $>$32 &  & 0.01(2) & 4.2(4) & $>$91 \\
kT$_{bb}$ (keV)       & 1.6(1) & & &  & 1.46(4) & & \\
N$_{bb}$ ($^{b}$)     & 6.5(4)\raisebox{.0ex}[2.5ex][.0ex]{$\cdot10^{-5}$} & & &  & 6.8(2)\raisebox{.0ex}[2.5ex][.0ex]{$\cdot10^{-5}$} & & \\
{\tt AGA} ($^{a}$)    & 0.010(4) & & &  & 0.031(2) & & \\
{\tt APL} ($^{a}$)    & 0.4(4) & 3.2(1) & 14.0(4) &  & 0.12(1) & 2.43(5) & 13.7(2) \\
{\tt f}               & 0.09$^{+0.46}_{-0.07}$ & 0.90(2) & 0.955(2) &  & 0.4(3) & 0.853(1) & 0.963(1) \\
$\Gamma_{pl}$         & 1.33(1) & & &  & 1.31(1) & & \\
N$_{pl}$ ($^{c}$)     & 0.0181(4) & & &  & 0.0177(1) & & \\
E$_{line}$ (keV)      & -- & -- & -- &  & 0.524(2) & & \\
$\sigma_{line}$ (keV) & -- & -- & -- &  & 0.103(2) & & \\
N$_{line}$ ($^{d}$)   & -- & -- & -- &  & 0.032(1) & & \\
\hline
\end{tabular}
\end{center}
\noindent $^a$ 10$^{22}$ atoms cm$^{-1}$ ; $^b$ $L_{39}/(D_{10})^2$, with $L_{39}$ the luminosity ($10^{39}$ ergs s$^{-1}$) and $D_{10}$ the source distance (10 kpc) \\
\noindent $^c$ photons\,keV$^{-1}$\,cm$^{-2}$\,s$^{-1}$ at 1 keV ; $^d$ total photons\,cm$^{-2}$\,s$^{-1}$ in the line \\

\end{table*}

Spectral fits were made with the models used by \citet{chbado1998}
and \citet{bohafe2001}. The first model was only used in conjunction
with the count-rate selected spectra (L1--L5), whereas the
\citet{bohafe2001} model was used for both types of selection (to
study the effect of selection method).

\subsubsection{\citet{chbado1998} model}

The \citet{chbado1998} model consists of a black body component from
the neutron star, plus Comptonized emission from an extended and
progressively covered accretion disc corona (ADC): {\tt ABB*BB +
APL*f*PL + AGA*(1-f)*PL}, where {\tt BB} and {\tt PL} are the black
body and power-law terms, {\tt AGA} the Galactic absorption term,
{\tt ABB} and {\tt APL} the variable black body and power-law
absorption terms, and {\tt f} the progressive covering fraction of
the ADC. Note that in this definition of the model {\tt ABB} and {\tt
APL} include Galactic absorption and that the two {\tt PL} components
represent the covered and uncovered parts of the same spectral
component. In all fits with this model the columns densities ({\tt
ABB}, {\tt AGA} and {\tt APL}) have a lower limit of $10^{20}$
atoms/cm$^2$, which is a factor of $\sim$10 lower than the average
galactic column density in the direction of \exo\ \citep{dilo1990}.
{\tt AGA} and the parameters for the black body and the power law
were treated as single parameters for all spectra.

\begin{figure*}[t]
\centerline{\includegraphics[width=17.5cm]{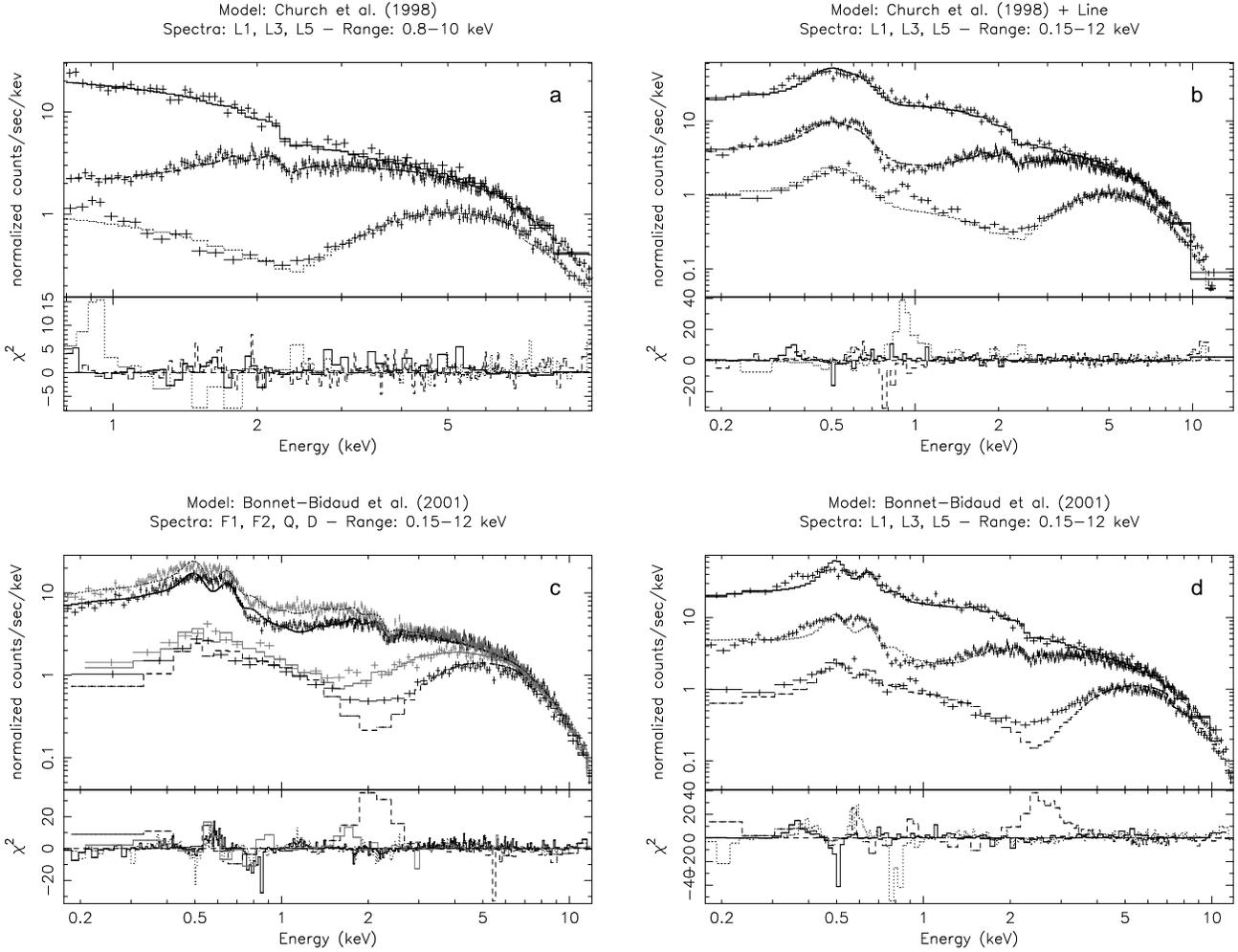}} 
\caption{Model fits to various \xmm\ spectra of \exo. Spectra,
models, and energy ranges are given above each panel. All spectra
were rebinned for display purposes. In the panels a, b, and d spectra
L5, L3 and L1 are plotted from plot to bottom. In panel c the order
of the spectra, from top to bottom, is F2, F1, Q and D. See Table
\ref{tab:fits} for fit parameters.} \label{fig:fits} \end{figure*}

We started by performing fits to the L1, L3 and L5 spectra in the
energy range used by \citet{chbado1998} (0.8--10 keV).
Fig.~\ref{fig:fits}a shows the results of this fit; the best-fit
parameters can be found in Table \ref{tab:fits-church}. The  
$\chi^2_{red}$/d.o.f of 1.01/1594 is good, although {\tt ABB} for
spectrum L3 seems to be too low, given the covering fraction of 90\%
that we find. More worrisome are the unacceptably large residuals
below 1 keV and also around 1.6 keV in the fit to spectrum L1. This
part of the spectrum is not shown in Fig.~3 of \citet{chbado1998}, so
we cannot check whether such residuals were also present in their
fits to the ASCA data. Most likely, these residuals are due to the
soft excess found by \citet{thcosm1997} and also discussed by
\citet{chbado1998}. The presence of the this soft component is much
clearer in the 0.15--12 keV spectra and a fit with the model in this
energy range is clearly worse than in the 0.8--10 keV band
($\chi^2_{red}$=1.76). Following \citet{chbado1998} we add a Gaussian
line to the model, which, like the power law component, is
progressively covered. The best fit with this model is shown in
Fig.~\ref{fig:fits}b; the best-fit parameters can be found in Table
\ref{tab:fits-church}. Again, even though the $\chi^2_{red}$=1.18 is
acceptable, large (localized) residuals are present - this time in
the fits to all three spectra. As a general remark we note, that
since most data bins are in the rather featureless part of the
spectrum above a few keV, which is fitted well by the models, the
residuals below 2 keV (which is the part of the spectrum we are most
interested in) have only a limited effect on the $\chi^2_{red}$.

When plotting the separate spectral components it becomes clear that
rather than being a narrow line, the Gaussian is very broad and fits
a large part of the continuum below 1 keV. Moreover, the ADC
component dominates the spectrum over the entire energy range in all
three spectra. The covering fraction and the two separate absorption
components ({\tt AGA} and {\tt APL}) act in such a way on the (very
hard) ADC component that the uncovered power law (together with the
Gaussian) fits the low energy part of the spectrum, whereas the
covered power law fits the high energy part. In the L5 spectrum the black body component only contributes 20\% of the absorbed 0.15-12 keV flux.

\begin{table*}
\caption{Spectral fit results. Models and energy ranges are given for
each of the fits. All errors represent 1$\sigma$ confidence
intervals.}\label{tab:fits}
\begin{center}
\begin{tabular}{lllllllll}
\hline \hline
 & \multicolumn{4}{l}{Model: Bonnet-Bidaud et al.\,(2001) } & & \multicolumn{3}{l}{Model: Bonnet-Bidaud et al.\,(2001)} \\
 & \multicolumn{4}{l}{Range: 0.15--12 keV \mbox{~~~} $\chi^2_{red}$/d.o.f.=1.35/2904} & & \multicolumn{3}{l}{Range: 0.15--12 keV \mbox{~~~}
$\chi^2_{red}$/d.o.f.=1.54/1914} \\
\hline
 & F1 & F2 & Q & D & & L5 & L3 & L1 \\
\hline
{\tt APL} ($^{a}$) & 1.82(2) & 1.28(2) & 5.7(1) & 11.6(2) &  & 1.34(6) & 2.77(3) & 17.9(3) \\
$\Gamma_{pl}$      & 1.29(1) & & & &  & 1.44(1)  & & \\
N$_{pl}$ ($^{b}$)  & 2.19(1)\raisebox{.0ex}[2.5ex][.0ex]{$\cdot10^{-2}$} & & & &  & 2.85(2)\raisebox{.0ex}[2.5ex][.0ex]{$\cdot10^{-2}$} & & \\
{\tt ARS} ($^{a}$) & 0.048(1)& 0.052(1) & 0.18(1) & 0.24(1) &  & 0.033(2) & 0.010(1) & 0.110(4) \\
kT$_{RS}$ (keV)    & 0.50(1) & & & &  & 1.42(3) & & \\
N$_{RS}$ ($^{d}$)  & 3.35(4)\raisebox{.0ex}[2.5ex][.0ex]{$\cdot10^{-2}$} & 4.86(5)\raisebox{.0ex}[2.5ex][.0ex]{$\cdot10^{-2}$} &
2.0(1)\raisebox{.0ex}[2.5ex][.0ex]{$\cdot10^{-2}$} & 1.98(6)\raisebox{.0ex}[2.5ex][.0ex]{$\cdot10^{-2}$} &  & 3.79(1)\raisebox{.0ex}[2.5ex][.0ex]{$\cdot10^{-2}$} &
5.6(1)\raisebox{.0ex}[2.5ex][.0ex]{$\cdot10^{-3}$} & 2.77(5)\raisebox{.0ex}[2.5ex][.0ex]{$\cdot10^{-3}$}\\
\hline
\end{tabular}
\end{center}
\noindent $^a$ 10$^{22}$ atoms cm$^{-1}$ ; $^b$ $L_{39}/(D_{10})^2$, with $L_{39}$ the luminosity ($10^{39}$ ergs s$^{-1}$) and $D_{10}$ the source distance (10 kpc) \\
\noindent $^c$ photons\,keV$^{-1}$\,cm$^{-2}$\,s$^{-1}$ at 1 keV ; $^d$ $10^{-14}/(4\pi D^2_A)\int n_e n_H dV$ , with where $D_A$ is
	   the angular size distance to the source (cm), $n_e$ is the electron
	   density (cm$^{-3}$), and $n_H$ is the hydrogen density (cm$^{-3}$) \\
\end{table*}
\subsubsection{\citet{bohafe2001} model}

In the \citet{bohafe2001} model the source consists of a spectrally
hard compact central component and a spectrally soft extended
component. These are modeled, respectively, by a power law ({\tt PL})
a thermal Raymond-Smith model ({\tt RS}), each affected by their own
absorption component: {\tt APL*PL + ARS*RS}, where {\tt APL} and {\tt
ARS} both include the galactic absorption and are not allowed to
become lower than $10^{20}$ atoms/cm$^2$.

The first fits with the \citet{bohafe2001} model were made to the
time selected 0.15--12.0 keV spectra. The element abundances were
fixed to the values found by \citet{bohafe2001}, while the other
parameters were allowed to vary. The power law index, the power law
normalization and the temperature of the thermal component were
treated as single parameters in the joint fit to all four spectra. We
obtain very similar values as \citet{bohafe2001}, but with a slightly
higher $\chi^2_{red}$ of 1.34 (d.o.f.=2904) compared to their
$\chi^2_{red}$/d.o.f.=1.27/2467. This difference can be partly
attributed to a different grouping of the data (minimum of 25 counts
per bin instead of 20), higher maximum energy (12 keV instead of 10
keV), and also due to improved calibration of the  instruments. In
any case, as is clear from Fig.~\ref{fig:fits}a, large localized
residuals and broad deviations are present; e.g. in the fit to the dip
spectrum around 2 keV, and in the fits to the flare and quiescent
spectra above 3 keV.

\citet{bohafe2001} do not argue why they did not treat the
normalization of the thermal component as a single parameter for all
four spectra, as they did for the normalization of the power law
component. It is not clear to us why the properties of an extended
component would vary on a shorter time scale than those of a compact
component. Their argument that the increases in the normalization of
the thermal component are due to reprocessing of energy from type I
X-ray bursts is not supported by our observations of bursts and
flaring intervals. The increases in the normalization occur with
decreases in the absorption  of the thermal component, suggesting
that the variable normalization tries to compensate for the absence
of a progressive covering factor in the model. Indeed, tying the
value of the normalization to a single parameter results in a
$\chi^2_{red}$ of 1.92. 

Next we fitted the \citet{bohafe2001} model (i.e.\ without tying the
normalization of the thermal component, to stay consistent with the
previous fit) to our intensity selected spectra L1, L3, and L5 with
the abundances of N, O, Ne, and Mg left free to vary. Using the
fit-results from the time selected spectra as a starting point we
obtained a $\chi^2_{red}\sim$1.6 with similar fit parameters.
However, a lower $\chi^2_{red}$ of 1.53 was obtained with a higher
temperature and a much higher abundance of N ($\sim$80 times solar
compared to $\sim$10 times); the fit results can be found in Table
\ref{tab:fits} and are shown in Fig.~\ref{fig:fits}d. Again large
residuals are present around 2--3 keV in the L1 spectrum. Unlike the
fit to the time selected spectra, here the normalization and the
absorption of the extended component do not correlate.

\subsubsection{Eclipse spectrum}

A powerful test of both spectral models can be made with the spectrum
obtained during the eclipse. As the compact component is eclipsed,
one only expects to see contribution from the extended component. In
the \citet{chbado1998} model this is the (progressively covered)
power law + Gaussian combination and in the \citet{bohafe2001} model
this is the thermal halo. A fit with a power law + Gaussian at 0.52
keV gives $\chi^2_{red}$/d.o.f.=1.5/18, whereas the fit with the
Raymond-Smith model gives $\chi^2_{red}$/d.o.f.=1.9/15. It should be
noted that for the latter model we obtain both very high temperatures
and abundances. Fixing these parameters to more reasonable values
(like those obtained from the fits to the non-eclipse spectra) gives
a $\chi^2_{red}$ values of 3.5-5, with large positive residuals above
2 keV.

Given the strong indications that the scattered/reflected light from
compact source/neutron star {\it does} contribute to the eclipse in
this source - as suggested by observations of type I X-ray bursts
during eclipse \citep{pawhgi1985,gohapa1987} - it is not surprising
that both models fail to fit the eclipse spectrum. We therefore
include the compact components again in both models, keeping in mind
that 1) their contribution should be around 5\% of that measured
outside eclipse (i.e. the percentage of scattered flux measured for
the type I X-ray bursts in eclipse) and 2) that the $N_H$ for this
component should not be larger than that during dips, as the
scattered light should be much less affected by the absorbing
structures. Adding the black body (with kT fixed to 1.46) to the
power law + Gaussian combination does not lead to a significant
improvement of the fit. However, adding a power law (with index fixed
to 1.35 and abundance all to 1) to the Raymond-Smith model leads to a
much improved fit: $\chi^2_{red}$/d.o.f.=1.12/19, with kT$\sim$0.4.
The normalization of the power law component in this fit is
$\sim$2-3\% of that in the fits to the non-eclipse data and the $N_H$
is $\sim8\cdot10^{21}$ atoms cm$^{-2}$, in reasonable agreement with
our expectations and also similar to the results obtained from
eclipse spectra in MXB 1659--298 \citep{sioopa2001}. Assuming that a
considerable part of the reflecting/scattering medium is obscured
during the eclipse, it is quite likely that the contribution to the
spectrum of this reflected/scattered component is considerably higher
outside of eclipse, partly explaining the problems encountered when
fitting those spectra.

\subsection{Concluding remarks}

Our fits clearly show that current models are not able to fit the
absorption affected spectra of \exo\ during dips. While the fits to
the eclipse spectrum favor the thermal halo model of
\citet{bohafe2001} this model has severe problems with the deep dip
spectra. Our eclipse spectrum provided strong evidence for the
presence of an additional scatter/reflection component, which should
be an integral part in future models. Such models should also include
more advanced models for the absorbing material and also for the
structure of the thermal halo.

\end{document}